\documentstyle[preprint,aps]{revtex}
\tighten
\draft
\widetext
\input epsf
\preprint{YITP-SB-01-41, NYU-TH/01/07/09}
\bigskip
\bigskip

\begin{document}
\title{A Remark on Non-conformal Non-supersymmetric Theories
 with Vanishing Vacuum Energy Density}
\medskip
\author{Olindo Corradini$^1$\footnote{E-mail: olindo@insti.physics.sunysb.edu},
Alberto Iglesias$^1$\footnote{E-mail: iglesias@insti.physics.sunysb.edu},
Zurab Kakushadze$^{1,2}$\footnote{E-mail: zurab@insti.physics.sunysb.edu} and
Peter Langfelder$^1$\footnote{E-mail: plangfel@insti.physics.sunysb.edu}}
\bigskip
\address{$^1$C.N. Yang Institute for Theoretical Physics\\ 
State University of New York, Stony Brook, NY 11794\\
$^2$Department of Physics, New York University, New York, NY 10003}

\date{July 19, 2001}
\bigskip
\medskip
\maketitle

\begin{abstract} 
{}We discuss non-conformal non-supersymmetric 
large $N$ gauge theories with vanishing
vacuum energy density to all orders in perturbation theory.
These gauge theories can be obtained via a field theory limit of Type IIB
D3-branes embedded in orbifolded space-times. We also discuss gravity in
this setup. 
\end{abstract}

\section{Introduction}

{}The cosmological constant problem is probably the most severe problem
of the ``naturalness and hierarchy'' type. Its smallness compared with the
energy scales believed to be ``more'' fundamental is puzzling. However, 
perhaps even more puzzling is the fact that generically in known
non-supersymmetric field theories loop corrections to the vacuum energy 
density are expected to be as large as $\sim M_{\rm {\small SUSY}}^4$, where 
$M_{\rm {\small SUSY}}$ is the supersymmetry breaking scale. 

{}In principle, when discussing the cosmological constant problem, one
must include gravity. This makes the problem even more complicated, 
in particular, to treat gravity quantum mechanically at present we must appeal
to string theory, where supersymmetry breaking is still poorly understood.
However, even if we treat gravity as non-dynamical, the problem is still 
non-trivial. Thus, imagine that we have an interacting 
renormalizable four-dimensional field theory with supersymmetry broken at 
some scale $M_{\rm {\small SUSY}}$, which can be consistently coupled to 
gravity. Suppose now we treat gravity purely classically (that is,
we choose an appropriate gravitational background, and ignore fluctuations
around it), and wish to compute
loop contributions to the vacuum energy density coming from diagrams where
only the field theory states run in the loops. Generically, that is, without
fine-tuning, we still expect these loop contributions\footnote{Note that if
there is a condensate in a field theory such as the Higgs condensate in the
Standard Model, we will have additional contributions to the vacuum energy
density $\sim M_*^4$, where $M_*$ is the scale of the condensate.} 
to be as large as $\sim M_{\rm {\small SUSY}}^4$, with the possible 
exception of conformal field theories, where such contributions could vanish.

{}The purpose of this note is to present examples of non-conformal 
non-supersymmetric gauge theories with vanishing
vacuum energy density to all orders in perturbation theory.
The theories we discuss here are not realistic as they are large $N$ gauge
theories. Nonetheless, at least to the best of our knowledge, they are the
first examples of the aforementioned type. The key fact that enables one to 
show that in these theories the vacuum energy density vanishes to all orders 
in perturbation theory is that these gauge theories can be obtained via a 
field theory limit of Type IIB D3-branes embedded in orbifolded space-times. 
One then can use the power of string perturbation techniques to prove a
non-renormalization theorem for the vacuum energy density. 

{}This is precisely where 't Hooft's large $N$ limit \cite{tHooft} 
becomes crucial. In this limit the gauge theory diagrams are
organized in terms of Riemann surfaces, where each extra handle on the 
surface suppresses the corresponding diagram by 
$1/N^2$. The large $N$ expansion,
therefore, resembles perturbative expansion in string theory. In the case
of four-dimensional gauge theories this connection can be made precise in
the context of type IIB string theory in the presence of a large number $N$
of D3-branes \cite{z1}\footnote{The generalization of this setup in the
presence of orientifold planes was discussed in \cite{z2}.}. 
Thus, we consider a limit where 
$\alpha'\rightarrow 0$, $g_s\rightarrow 0$ and $N\rightarrow \infty$, 
while keeping $\lambda \equiv N g_s$ fixed, where $g_s$ is the type IIB
string coupling. Note that in this context a world-sheet with $g$
handles and $b$ boundaries is weighted with
\begin{equation}
 (N g_s)^b g_s^{2g-2}=\lambda^{2g-2+b} N^{-2g+2}~.
\end{equation}
Once we identify $g_s=g_{\rm{\small YM}}^2$, this is the same as the large 
$N$ expansion considered by 't Hooft. 
Note that for this expansion to make sense we must
keep $\lambda$ at a small value $\lambda <1$.
In this regime we can map the string diagrams directly to
(various sums of) large $N$ Feynman diagrams.
Note, in particular, that the genus $g=0$ planar diagrams
dominate in the large $N$ limit\footnote{Note that if $\lambda >1$, then
no matter how large $N$ is, for sufficiently many boundaries the
higher genus terms become relevant, and we lose the genus expansion. In fact,
in this regime one expects
an effective supergravity description to take over as discussed in
\cite{malda,GKP,Witten}.}.

{}If the space transverse to the D3-branes in the setup of \cite{z1} is 
${\bf R}^6$, then we obtain the ${\cal N}=4$ supersymmetric $U(N)$ gauge theory
on the D3-branes, which is conformal. On the other hand, we can also consider
orbifolds of ${\bf R}^6$, which leads to gauge theories with reduced 
supersymmetry. As was shown in \cite{z1}, if we cancel all twisted tadpoles
in such models, in the large $N$ limit the corresponding ${\cal N}=0,1,2$
gauge theories are conformal\footnote{The $\lambda>1$ versions of these 
orbifold theories
via the compactifications of type IIB on AdS$_5\times ({\bf S}^5/\Gamma)$
(where $\Gamma$ is the orbifold group) were originally discussed in 
\cite{Kachru}.}. Moreover, in the planar limit 
the (on-shell) correlation functions
in such theories are the same as in the parent ${\cal N}=4$ gauge theory
\cite{z1}.   

{}Recently, a generalization of the setup of \cite{z1} was discussed in
\cite{radu}, which allows one to obtain non-conformal gauge theories
by allowing some twisted tadpoles to be 
non-vanishing. In particular, we can have consistent embeddings
of non-conformal gauge theories if we allow logarithmic tadpoles, 
which correspond to the twisted sectors with fixed point loci of real dimension
two. Thus, even though the corresponding string backgrounds are
not finite (in the sense that we have logarithmic ultra-violet divergences),
they are still consistent as far as the gauge theories are concerned, 
and the divergences correspond to the running in the
four-dimensional gauge theories on the D3-branes.

{}Using the setup of \cite{radu}, we can construct non-conformal 
non-supersymmetric large $N$ gauge theories with vanishing 
vacuum energy density
to all orders in perturbation theory. 
In fact, in the planar limit the (on-shell) correlation functions in these 
theories are the same as in the parent ${\cal N}=2$ or ${\cal N}=1$
supersymmetric non-conformal gauge theories. In the former case these
gauge theories are not renormalized beyond one loop. 

{}The remainder of this paper is organized as follows. In section II we 
review the setup of \cite{radu}.
In section III we discuss non-conformal non-supersymmetric
large $N$ gauge theories
which can be constructed within this setup, and prove the non-renormalization
theorem for the vacuum energy density. Section IV contains some concluding
remarks, in particular, we discuss gravity in this setup.

\section{Setup}

{}In this section we review the setup of \cite{radu}. 
Thus, consider type IIB string theory in the presence of $N$ coincident  
D3-branes with the space transverse to the D-branes 
${\cal M}={\bf R}^6/\Gamma$. The orbifold group 
$\Gamma=\{g_a|a=1,\dots,|\Gamma|\}$ ($g_1=1$) must be 
a finite discrete subgroup of $Spin(6)$. If $\Gamma\subset SU(3)(SU(2))$, 
we have ${\cal N}=1$ (${\cal N}=2$) unbroken supersymmetry, and ${\cal N}=0$,
otherwise.

{}We will confine our attention to the cases where type IIB on ${\cal M}$ is
a modular invariant theory\footnote{This is always the case if there are
some unbroken supersymmetries. If all supersymmetries are broken, this is also
true if $\not\exists{\bf Z}_2\subset\Gamma$. If 
$\exists{\bf Z}_2\subset\Gamma$, then modular invariance requires 
that the set of points in ${\bf R}^6$
fixed under the ${\bf Z}_2$ twist has real dimension 2.}. The action of the
orbifold on
the coordinates $X_i$ ($i=1,\dots,6$) on ${\cal M}$ can be described
in terms of $SO(6)$ matrices:
$g_a:X_i\rightarrow (g_a)_{ij} X_j$.
The world-sheet fermionic superpartners of $X_i$
transform in the same way.
We also need to specify
the action of the orbifold group on the Chan-Paton charges carried by the
D3-branes. It is described by $N\times N$ matrices $\gamma_a$ that
form a representation of $\Gamma$. Note that $\gamma_1$ is an
identity matrix, and ${\rm Tr}(\gamma_1)=N$.

{}The D-brane sector of the theory is described by an oriented open
string theory. In particular, the world-sheet expansion corresponds
to summing over oriented Riemann surfaces with arbitrary genus $g$ and
arbitrary
number of boundaries $b$, where the boundaries of the world-sheet are mapped
to the D3-brane world-volume. Moreover, we must consider various ``twists''
around the cycles of the Riemann surface. The choice of these ``twists''
corresponds to a choice of homomorphism of the fundamental group of
the Riemann surface with boundaries to $\Gamma$.

{}For example, consider the one-loop vacuum amplitude ($g=0, b=2$). The
corresponding graph is an annulus whose boundaries lie on D3-branes.
The annulus amplitude is given by
\begin{equation}
 {\cal C}=\int_0^\infty {dt\over t} ~{\cal Z}~.
\end{equation}
The one-loop partition function 
${\cal Z}$ in the light-cone gauge is given by
\begin{equation}\label{partition}
 {\cal Z}={1\over |\Gamma|}\sum_a
 {\rm Tr}  \left( g_a ~{{1-(-1)^F}\over 2}~
 e^{-2\pi tL_0}
 \right)~,
\end{equation}
where $L_0$ is the light-cone Hamiltonian,
$F$ is the fermion number operator, $t$ is the real modular parameter
on the cylinder, and the trace includes sum over the Chan-Paton factors.
The states in the Neveu-Schwarz (NS) sector are space-time bosons and enter the
partition function with weight $+1$, whereas the states in the Ramond (R)
sector are space-time fermions and contribute with weight $-1$.

{}The elements $g_a$ acting in the Hilbert space of open strings
act both on the left end and the right end of the open string.
This action
corresponds to $\gamma_a\otimes \gamma_a$ acting on the Chan-Paton indices.
The partition function (\ref{partition}), therefore,
has the following form:
\begin{equation}\label{gsq}
 {\cal Z}={1\over |\Gamma|}
 \sum_a \left({\rm Tr}(\gamma_a)\right)^2 {\cal Z}_a~,
\end{equation}
where ${\cal Z}_a$ are characters
corresponding to the world-sheet degrees of freedom. The ``untwisted''
character
${\cal Z}_1$ is the same as in the ${\cal N}=4$ theory for which
$\Gamma=\{1\}$. The information about the fact that the orbifold theory
has reduced supersymmetry is encoded in the ``twisted'' characters
${\cal Z}_a$, $a\not=1$.

\subsection{Tadpole Cancellation}

{}In this subsection we discuss one-loop tadpoles arising in the above 
setup. As was pointed out in \cite{z1}, if all tadpoles are canceled, then
the resulting theory is finite in the large $N$ limit. However, not all 
tadpoles need to be canceled to have a consistent four-dimensional gauge 
theory. In fact, we can obtain non-conformal gauge theories if we allow such
tadpoles.

{}The characters ${\cal Z}_a$ in (\ref{gsq}) are given by
\begin{equation}
 {\cal Z}_a={1\over (8\pi^2\alpha^\prime t)^2 }{1\over 
 \left[\eta(e^{-2\pi t})\right]^{2+d_a}}
 \left[{\cal X}_a(e^{-2\pi t})-{\cal Y}_a(e^{-2\pi t})\right]~,
\end{equation}
where $d_a$ is the real dimension of the set of
points in ${\bf R}^6$ fixed under the twist $g_a$. The factor of $(8\pi^2
\alpha^\prime t)^2$ in the denominator comes from the bosonic zero modes 
corresponding to four directions along the D3-brane 
world-volume. Two of the $\eta$-functions come from the
oscillators corresponding to two spatial directions along the D3-brane 
world-volume (the time-like and longitudinal contributions are
absent as we are working in the light-cone gauge). The other $d_a$
$\eta$-functions come from the oscillators corresponding
to the directions transverse to the D-branes untouched by the orbifold
twist $g_a$. Finally, the
characters ${\cal X}_a$, ${\cal Y}_a$ correspond to the contributions of
the world-sheet fermions, as well as the
world-sheet bosons with $g_a$ acting non-trivially
on them (for $a\not=1$): 
\begin{eqnarray}
 &&{\cal X}_a ={1\over 2}{\rm Tr}^\prime \left[g_a e^{-2\pi t L_0}\right]~,\\
 &&{\cal Y}_a ={1\over 2}
 {\rm Tr}^\prime \left[g_a (-1)^F e^{-2\pi t L_0}\right]~,
\end{eqnarray}
where prime in ${\rm Tr}^\prime$ indicates that the trace is restricted as
described above.

{}For the annulus amplitude we therefore have
\begin{equation}
 {\cal C}={1\over (8\pi^2\alpha^\prime)^2 }{1\over |\Gamma|} \sum_a 
 \left[{\cal A}_a - 
 {\cal B}_a\right]~,
\end{equation}
where
\begin{eqnarray}
 &&{\cal A}_a=\left({\rm Tr}\left(\gamma_a\right)\right)^2 
 \int_0^\infty {dt\over t^3}
 {1\over \left[\eta(e^{-2\pi t})\right]^{2+d_a}}{\cal X}_a(e^{-2\pi t})~,\\
 &&{\cal B}_a=\left({\rm Tr}\left(\gamma_a\right)\right)^2 
 \int_0^\infty {dt\over t^3}
 {1\over \left[\eta(e^{-2\pi t})\right]^{2+d_a}}{\cal Y}_a(e^{-2\pi t})~.
\end{eqnarray}
These integrals\footnote{For space-time supersymmetric
theories the total tadpoles vanish: ${\cal A}_a-{\cal B}_a=0$. (The entire
partition function vanishes as the numbers of
space-time bosons and fermions are equal.) For consistency, however, we must
extract tadpoles from individual contributions ${\cal A}_a$ and ${\cal B}_a$.
Thus, for instance, cancellation of certain tadpoles coming from ${\cal B}_a$
is required for consistency of the equations of motion
for the twisted R-R four-form which couples to D3-branes (see below).}
are generically divergent as $t\rightarrow 0$ reflecting the 
presence of tadpoles.
To extract these divergences we can change variables $t=1/\ell$ so that the
divergences correspond to $\ell\rightarrow\infty$:
\begin{eqnarray}\label{NSNS}
 &&{\cal A}_a =\left({\rm Tr}(\gamma_a)\right)^2 \int_0^\infty
 {d\ell\over \ell^{d_a/2}}
  \sum_{\sigma_a} N_{\sigma_a} e^{-2\pi\ell \sigma_a}~,\\
 \label{RR}
 &&{\cal B}_a =\left({\rm Tr}(\gamma_a)\right)^2 \int_0^\infty
  {d\ell\over \ell^{d_a/2}}
 \sum_{\rho_a} N_{\rho_a}e^{-2\pi\ell \rho_a}~.
\end{eqnarray}
The closed string states contributing to ${\cal A}_a$ (${\cal B}_a$)
in the transverse channel
are the NS-NS (R-R) states with $L_0={\overline L}_0=\sigma_a (\rho_a)$
(and $N_{\sigma_a} (N_{\rho_a})>0$ is
the number of such states). The massive states with
$\sigma_a(\rho_a)>0$
do not lead to divergences as $\ell\rightarrow\infty$.
On the other hand, the divergence property of the above integrals
in the $\ell\rightarrow\infty$ limit is determined by the value of $d_a$.
Given the orientability of $\Gamma$ the allowed values of 
$d_a$ are $0,2,4,6$. For $d_1=6$ there is no divergence in ${\cal B}_1$, so
we have no restriction for ${\rm Tr}(\gamma_1)=N$. For $d_a=4$ the
corresponding twisted NS-NS closed string sector contains tachyons. 
This leads to a tachyonic divergence in ${\cal A}_a$ unless 
${\rm Tr}(\gamma_a)=0$ for the corresponding $g_a$ twisted sector.
Next, if a twist $g_a$ with $d_a=2$ ($d_a=0$) is non-supersymmetric, 
that is, if $g_a \in Spin(6)$ but $g_a\not\in SU(2)$ ($g_a\not\in SU(3)$), then
we have tachyons in the corresponding NS-NS twisted sector,
so we must require ${\rm Tr}(\gamma_a)=0$ to avoid a tachyonic divergence.
Finally, if a twist $g_a$ with $d_a=2$ ($d_a=0$) is supersymmetric,
that is, if $g_a\in SU(2)$ ($g_a\in SU(3)$), then we have massless states
in the corresponding R-R as well as NS-NS twisted sectors, so we 
get divergences due to massless 
R-R states in the integral in ${\cal B}_a$ and due to massless NS-NS states
in the integral in ${\cal A}_a$ for large $\ell$
for such twists with $d_a=0,2$. 

{}We must therefore consider massless tadpoles
arising for supersymmetric twists with
$d_a=0,2$. For $d_a=0$ the corresponding integrals are
linearly divergent with $\ell$ as $\ell\rightarrow\infty$. To cancel
such a tadpole we must require that ${\rm Tr}\left(\gamma_a\right)=0$
for the corresponding $g_a$ twisted sector. On the other hand, if such
a tadpole is not canceled, in the four-dimensional field theory language this 
would correspond to having a {\em quadratic} (in the momentum) divergence at 
the one-loop order. This would imply that the corresponding four-dimensional
background is actually inconsistent, in particular, the equation of motion for
the corresponding R-R twisted four-form (which couples to the D3-brane 
world-volume) is inconsistent. Thus, we must require that for such twists  
${\rm Tr}(\gamma_a)=0$.

{}Finally, let us discuss supersymmetric twists with 
$d_a=2$. For such twists the corresponding
integrals are only logarithmically divergent as $\ell\rightarrow\infty$.
If such a tadpole is not canceled, that is, if the corresponding
${\rm Tr}(\gamma_a)\not=0$, in the four-dimensional field theory language 
this corresponds to having a {\em logarithmic} divergence (in the momentum)
at the one-loop order. As was pointed out in \cite{radu}, these logarithmic 
divergences are precisely related to the running in the corresponding gauge
theories, which are {\em not} conformal (even in the large $N$ limit).
Note that the corresponding twisted closed string states propagate in
two extra dimensions transverse to the D3-branes (which correspond to the
locus of the points fixed under the twist $g_a$), 
so that the four-dimensional backgrounds
are perfectly consistent - the tadpoles for these fields simply imply that 
these fields have non-trivial (logarithmic)
profiles in these two extra dimensions (while
the four dimensions along the D-brane world-volume are flat)\footnote{In fact,
as was shown in \cite{radu} for general $\Gamma$, the presence of 
such tadpoles does not introduce any 
anomalies. This was shown for $\Gamma\approx {\bf Z}_m\otimes 
{\bf Z}_n$ in \cite{rozleigh} in a somewhat more 
complicated way.}. Thus, we conclude that, to obtain consistent four 
dimensional gauge theories in the D3-brane world-volume, we must require that
all twisted ${\rm Tr}(\gamma_a)=0$ except for supersymmetric twists with
$d_a=2$. If ${\rm Tr}(\gamma_a)=0$ for all 
such twists as well, then in the large
$N$ limit we get conformal theories \cite{z1}. On the other hand, if 
any of such twisted $\gamma_a$ is not traceless, the corresponding theories 
are not conformal even in the large $N$ limit \cite{radu}.

\section{Non-Supersymmetric Theories}

{}In this section we discuss non-supersymmetric large $N$ gauge theories 
arising in the above 
setup with some supersymmetric 
twists $g_a$ with $d_a=2$ such that ${\rm Tr}(\gamma_a)\not=0$.
As we have already mentioned, such theories are not conformal.
In these theories the gauge group 
is a product of $U(N_k)$ factors, and we also have matter, which
can be obtained using the corresponding quiver diagrams 
(see \cite{dougmoore,LNV}\footnote{A T-dual description of such models
can be studied in the context of the brane-box models \cite{HSU}.}). 
There is
always an overall center-of-mass $U(1)$, which is free. Other $U(1)$ factors,
however, run as the matter is charged under them. In the large $N$ limit,
however, these $U(1)$'s decouple in the infra-red, and can therefore be 
ignored. More precisely, in some cases we have
anomalous $U(1)$'s. Thus, in cases where we have twists with 
$d_a=0$ some of the 
$U(1)$ factors are actually anomalous (in particular,
we have mixed $U(1)_k SU(N_l)^2$ anomalies), 
and are broken at the tree-level via
the generalized Green-Schwarz mechanism \cite{IRU,Poppitz}. 
In other cases, where we have 
non-anomalous $U(1)$'s, the latter decouple in the infra-red
in the large $N$ limit. At any rate, all of the $U(1)$ factors can be ignored
in the large $N$ limit, so that we can focus on the non-Abelian part of the
gauge group.

{}To obtain such models, consider an orbifold group $\Gamma$, which is a 
subgroup of $Spin(6)$ but is not a subgroup of $SU(3)$. 
Let ${\widetilde \Gamma}$
be a non-trivial subgroup of $\Gamma$ such that ${\widetilde \Gamma}\subset
SU(2)$. We will allow the Chan-Paton matrices $\gamma_a$ corresponding to the
non-trivial twists $g_a\in {\widetilde \Gamma}$ (which have $d_a=2$) not 
to be traceless, so that the corresponding
${\cal N}=2$ model is not conformal. However, we will require that
the other Chan-Paton matrices
$\gamma_a$ for the twists $g_a\not\in{\widetilde \Gamma}$ be traceless.
The resulting ${\cal N}=0$ model is not conformal. 
However, as was shown in \cite{radu}, in the planar
limit the on-shell correlation functions in the 
${\cal N}=0$ gauge theory are the same as in the parent ${\cal N}=2$
gauge theory corresponding to the orbifold group ${\widetilde \Gamma}$.
That is, in the large $N$ limit the 
perturbative ${\cal N}=0$ gauge theory 
amplitudes are not renormalized beyond one loop
(as usual, various running $U(1)$'s decouple in the IR in
this limit). 

{}Note that 
the large $N$ property is crucial
here. The reason is that in the cases where the orbifold group $\Gamma\not
\subset SU(3)$, we always have twisted NS-NS 
closed string sectors with tachyons. Their contributions to the
corresponding part of the annulus amplitude (\ref{NSNS}) is then exponentially 
divergent unless we require that 
\begin{equation}
 {\rm Tr}\left(\gamma_a\right)=0~,~~~g_a\not\in SU(3)~.
\end{equation}
However, 
even if this condition is satisfied, we must take the 't Hooft limit - indeed,
otherwise it is unclear, for instance, how to deal with the diagrams with 
handles, which contain tachyonic divergences. In fact, the same applies to
some non-planar diagrams without handles, that is, diagrams where the 
external lines are attached to more than one boundaries (such diagrams are
subleading in the large $N$ limit).

\subsection{Vanishing of the Vacuum Energy Density}

{}As we already mentioned, in the planar
limit the on-shell correlation functions in the 
${\cal N}=0$ gauge theories of the aforementioned type 
are the same as in the corresponding parent ${\cal N}=2$ gauge theories.
In particular, this applies to zero-point functions corresponding to the
perturbative contributions to the vacuum energy density. That is, even though
these gauge theories are non-supersymmetric, the vacuum energy density vanishes
to all order in perturbation theory in such models.

The proof of this statement is straightforward. Thus, consider a vacuum
amplitude with $b$ boundaries but no 
handles (such a diagram corresponds to a $(b-1)$-loop diagram in the
field theory language). 
Next, we need to specify the twists on the boundaries. 
A convenient choice (consistent with that made for the annulus
amplitude (\ref{partition})) is given 
by\footnote{Here some care is needed in the 
cases where the orbifold group $\Gamma$ is non-Abelian, and we have to choose
base points on the world-sheet to define the twists. Our discussion here, 
however, is unmodified also in this case.}
\begin{equation}\label{mono}
 \gamma_{a_1}=\prod_{s=2}^{b} \gamma_{a_s}~,
\end{equation}
where $\gamma_{a_1}$ corresponds to the outer boundary, while 
$\gamma_{a_s}$, $s=2,\dots b$, correspond to inner boundaries.
Then the above vacuum amplitude has the following
Chan-Paton group-theoretic dependence:
\begin{equation}
 \sum
\prod_{s=1}^{b}  {\rm Tr}(\gamma_{a_s})~,
\end{equation}
where the sum involves all possible distributions of the $\gamma_{a_s}$ twists
that satisfy the condition (\ref{mono}). Note that the 
diagrams with all twists $g_a\in{\widetilde \Gamma}$ are (up to overall
numerical coefficients) the same as in the parent ${\cal N}=2$ theory, and 
therefore vanish. All other diagrams contain at least one
twist $g_{a_s}\not\in {\widetilde \Gamma}$.
This then implies that all such diagrams vanish as
\begin{equation}
 {\rm Tr}(\gamma_a)=0~,~~~g_a\not\in{\widetilde \Gamma}~.
\end{equation}
That is, in the large $N$ limit the vacuum energy density vanishes in such 
theories to all orders in perturbation theory. In fact, 
as was shown in \cite{radu}, in these theories on-shell correlation
functions are not renormalized beyond one loop. For instance,  
the non-Abelian gauge couplings do not run
in the large $N$ limit beyond one loop\footnote{More precisely, the
higher loop contributions to the gauge coupling running,
which come from the diagrams with handles, are subleading in the large $N$
limit compared with the leading one-loop contribution. This is analogous to 
what happens in theories discussed in \cite{KT}. In fact, the techniques
used in \cite{KT} 
to prove that the higher loop corrections are subleading are very
similar to the ones used in \cite{radu}.}.
 
\subsection{A Generalization}

{}Above we constructed ${\cal N}=0$ non-conformal gauge theories that have
parent ${\cal N}=2$ gauge theories. As far as vanishing of the vacuum energy 
density to all loop orders is concerned, 
it actually suffices that the parent theories are ${\cal N}=1$
supersymmetric (albeit in such cases the non-renormalization property
beyond one loop
for higher point functions is lost). Such theories can be 
obtained as follows. Thus, consider an orbifold group $\Gamma$, which is a 
subgroup of $Spin(6)$ but is not a subgroup of $SU(3)$. 
Let ${\widetilde \Gamma}$
be a non-trivial subgroup of $\Gamma$ such that ${\widetilde \Gamma}\subset
SO(3)$, but ${\widetilde \Gamma}\not\subset
SU(2)$. Note that all the non-trivial elements $g_a\in{\widetilde\Gamma}$
have $d_a=2$. We will therefore allow the corresponding 
Chan-Paton matrices $\gamma_a$ not 
to be traceless, so that the corresponding
${\cal N}=1$ model is not conformal. However, we will require that
the other Chan-Paton matrices
$\gamma_a$ for the twists $g_a\not\in{\widetilde \Gamma}$ be traceless.
The resulting ${\cal N}=0$ model is not conformal. 
In the planar
limit the on-shell correlation functions in the 
${\cal N}=0$ gauge theory are the same as in the parent ${\cal N}=1$
gauge theory corresponding to the orbifold group ${\widetilde \Gamma}$.
In particular, the vacuum energy density vanishes to all orders in 
perturbation theory.

\subsection{Examples}

{}Let us consider a simple example of such a theory. Let $\Gamma\approx {\bf
Z}_2\otimes {\bf Z}_3$, where the action of the generators $R$ and $\theta$ of
the ${\bf Z}_2$ respectively ${\bf Z}_3$ subgroups on the complex coordinates
$z_\alpha$ on ${\cal M}={\bf C}^3/\Gamma$ is as follows: $R:z_1\rightarrow
z_1$, $R:z_{2,3}\rightarrow -z_{2,3}$, $\theta:z_1\rightarrow \omega z_1$, 
$\theta: z_{2,3}\rightarrow z_{2,3}$, where $\omega\equiv\exp(2\pi i/3)$. The
twisted Chan-Paton matrices are given by: $\gamma_R=I_{3N}$, $\gamma_\theta=
{\rm diag}(I_N,\omega I_N,\omega^{-1} I_N)$. (Note that in this case
${\widetilde \Gamma}\approx {\bf Z}_2\subset SU(2)$.) Then the theory is a 
non-supersymmetric $SU(N)\otimes SU(N)\otimes SU(N)$ gauge theory 
(we are dropping the $U(1)$ factors for the reasons discussed above) with
matter consisting of complex scalars in $({\bf N},{\overline{\bf N}},{\bf 1})$,
$({\bf 1},{\bf N},{\overline{\bf N}})$ and 
$({\overline{\bf N}},{\bf 1},{\bf N})$, as well as chiral fermions in the
above representations plus their complex conjugates. Note that
the numbers of the physical bosonic and fermionic degrees of freedom
are the same. In fact, this is the case for all such gauge
theories.

{}Note that in the above example we have a {\em non-chiral} non-supersymmetric
gauge theory. Moreover, this theory contains massless scalars in the 
bifundamental representations. We can, however, construct examples of
{\em chiral} non-supersymmetric gauge theories. Moreover, such theories need 
not contain scalars. Thus, let $\Gamma\approx {\bf Z}_2\otimes {\bf
Z}_2\otimes {\bf Z}_3$, where the action of the generators $R_1$, $R_2$ 
and $\theta$ of
the first ${\bf Z}_2$, second ${\bf Z}_2$ and ${\bf Z}_3$ subgroups, 
respectively, on the complex coordinates
$z_\alpha$ on ${\cal M}={\bf C}^3/\Gamma$ is as follows: $R_1:z_1\rightarrow
z_1$, $R_1:z_{2,3}\rightarrow -z_{2,3}$, $R_2:z_2\rightarrow z_2$,
$R_2:z_{1,3}\rightarrow -z_{1,3}$,
$\theta:z_1\rightarrow \omega z_1$, 
$\theta: z_{2,3}\rightarrow z_{2,3}$, where $\omega\equiv\exp(2\pi i/3)$. The
twisted Chan-Paton matrices are given by: $\gamma_{R_1}=
\gamma_{R_2}=I_{3N}$, $\gamma_\theta=
{\rm diag}(I_N,\omega I_N,\omega^{-1} I_N)$. (Note that in this case
${\widetilde \Gamma}\approx{\bf Z}_2\otimes {\bf Z}_2\subset SO(3)$, but
${\widetilde \Gamma}\not\subset SU(2)$.)
Then the theory is a 
non-supersymmetric $SU(N)\otimes SU(N)\otimes SU(N)$ gauge theory 
(once again, we are dropping the $U(1)$ factors) with
matter consisting of chiral fermions $\Psi_1,\Psi_2,\Psi_3$
in $({\bf N},{\overline{\bf N}},{\bf 1})$,
$({\bf 1},{\bf N},{\overline{\bf N}})$ and 
$({\overline{\bf N}},{\bf 1},{\bf N})$, respectively. Since we have no scalars,
even classically there is no moduli space in this theory, so it corresponds to
an isolated non-supersymmetric vacuum. Moreover, this theory is chiral.
Also note that the operator
\begin{equation}
 B\equiv \Psi_1\Psi_2\Psi_3
\end{equation}
corresponds to a baryon of this theory.

\section{Remarks}

{}Thus, as we see, we can construct an infinite number of non-supersymmetric
non-conformal large $N$ gauge theories with vanishing vacuum energy density
to all orders in perturbation theory. In such a gauge theory the gauge group
is a product of $SU(N_k)$ factors, and we have charged matter. Here we would
like to comment on gravity in this setup.

{}Note that the Type IIB setup within which we discussed these gauge theories
can be thought of as a brane world scenario
\cite{early,BK,pol,witt,lyk,shif,TeV,dien,3gen,anto,ST,BW,Gog,RS,DGP,DG,IK}. 
In particular, in this case we have infinite-volume extra dimensions 
\cite{DGP,DG}. Since the brane matter is not conformal, we expect the
Einstein-Hilbert term to be generated in the D3-brane world-volume already at
the one-loop order \cite{DGP,DG}. In fact, the corresponding induced 
four-dimensional Planck scale
\begin{equation}
 M_P^2\sim N^2 \Lambda^2~,
\end{equation}
where $\Lambda$ is the gauge theory cut-off\footnote{One might wish to identify
$\Lambda$ with the string scale $M_s$. Note, however, that the string 
backgrounds we are considering here are not finite - we need an ultra-violet
cut-off to regularize logarithmic divergences (see \cite{radu} for details),
so identifying $\Lambda$ with $M_s$ is not necessary.}. The factor $N^2$
comes from the fact that the number of brane matter degrees of freedom 
propagating in the loops is of order $N^2$. 

{}Now, in the large $N$ limit $M_P$ goes to infinity, so, not surprisingly, 
we have no gravity on the branes. In the case of non-supersymmetric gauge 
theories we are essentially forced to consider the large $N$ limit to avoid
problems with bulk tachyons. That is, in this context we do not get 
four-dimensional gravity unless we are able to consider finite $N$ gauge
theories. Since the corresponding non-supersymmetric gauge theories are
perfectly consistent\footnote{Note that the non-Abelian gauge theories we are
discussing here are actually asymptotically free if
we choose the Chan-Paton matrices $\gamma_a$ corresponding to the 
elements $g_a$ of the orbifold subgroup ${\widetilde \Gamma}$
of the parent theory such that if ${\rm Tr}(\gamma_a)\not=0$
then $\gamma_a$ is an identity matrix.}, one might wish to argue
that the tachyon problem in their embedding in the Type IIB string theory
context might be an artifact of sorts. If so, then
one might hope to make sense of such embeddings for finite $N$
(perhaps a way to make this precise is
to consider $\alpha^\prime\rightarrow i\epsilon$ \cite{z1}). Note, however, 
that for finite $N$ we would lose any control over the vacuum energy density
as the arguments of the previous section crucially depend on the large $N$
property.

{}One tempting possibility around this difficulty is to start with a
supersymmetric gauge theory. In this case we can consider finite $N$ gauge
theories. A rosy scenario then goes as follows. Suppose the gauge theory on the
branes is actually ${\cal N}=1$ supersymmetric. Moreover, suppose supersymmetry
is dynamically broken on the branes via non-perturbative gauge dynamics. Then,
since the volume of the extra dimensions is infinite, bulk supersymmetry
is intact even if brane supersymmetry is completely broken \cite{DGP1,witten}.
Then in some cases unbroken bulk supersymmetry might protect the brane
cosmological constant \cite{DGP1,witten,zura,dvali}\footnote{Here we note that,
since the volume of the extra dimensions is infinite, 
the issues discussed in \cite{COSM}, which arise in scenarios with
finite-volume non-compact extra dimensions, need not concern us here.}. This
way we might hope to obtain a scenario where the brane cosmological constant
vanishes even though the brane supersymmetry is completely broken. Moreover,
the induced four-dimensional Planck scale on the branes in this case would be
finite as we are dealing with a finite $N$ gauge theory.

{}However, the above scenario seems to run into the usual
problem of runaway moduli. Indeed, for dynamical 
supersymmetry breaking to take place we need quantum modification of the 
moduli space. The latter does occur in some non-conformal ${\cal N}=1$ 
supersymmetric models of the type we are discussing here. In fact, a general
model of this type can be constructed as follows. Let the orbifold group
$\Gamma$ be a subgroup of $SU(3)$ but not a subgroup of $SU(2)$. Let 
${\widetilde \Gamma}$ be a non-trivial subgroup of $\Gamma$ such that 
${\widetilde \Gamma}\subset SO(3)$. Note that all the non-trivial
elements $g_a\in 
{\widetilde \Gamma}$ have $d_a=2$. We can therefore have ${\rm Tr}(\gamma_a)
\not=0$ for such elements $g_a\in {\widetilde \Gamma}$. For all the elements
$g_a\in\Gamma$ such that $g_a\not\in{\widetilde \Gamma}$ we will, however,
require ${\rm Tr}(\gamma_a) =0$. The resulting ${\cal N}=1$ gauge theory is
then non-conformal. Note that if ${\widetilde\Gamma}\not\subset SU(2)$, then
we do not have a parent ${\cal N}=2$ theory, and the ${\cal N}=1$ theory is
``truly'' ${\cal N}=1$ supersymmetric in the sense that even in the large 
$N$ limit the on-shell correlation functions are {\em not} the same as in an
${\cal N}=2$ supersymmetric model. On the other hand, if ${\widetilde\Gamma}
\subset SU(2)$, then we have a parent ${\cal N}=2$ theory, so that in the
large $N$ limit the on-shell correlation functions are the same as in
the parent ${\cal N}=2$ theory. As was pointed out in \cite{radu}, in some
${\cal N}=1$ 
theories of the latter type we have quantum modification of the moduli
space (such a modification is also present in some of the theories of the 
former type).

{}Albeit we can have quantum modification of the moduli space, supersymmetry
in these models is actually not broken as the twisted moduli that control the
corresponding gauge couplings have the usual runaway behavior. Moreover, 
tree-level couplings required for stabilization of these moduli via the
mechanism of \cite{dilaton} are also absent as the corresponding singlets
would have to come from the closed string sector which is ${\cal N}=2$
supersymmetric\footnote{In principle, we can generalize the above
construction to include orientifold planes, in which case the
closed string sector is ${\cal N}=1$ supersymmetric. 
However, in the presence of 
orientifold planes some caution is needed due to various issues discussed
in \cite{KaSh,KST,NPO}. 
At any rate, the required tree-level couplings would still
be absent in this case unless the corresponding singlets
come from twisted open string sectors
that arise in the context of non-perturbative orientifolds \cite{NPO}.}.

\acknowledgments

{}We would like to thank Gia Dvali and Radu Roiban for valuable discussions. 
This work was supported in part by the National Science Foundation and
an Alfred P. Sloan Fellowship.
Parts of this work were completed during Z.K.'s visit at New York University.
Z.K. would also like to thank Albert and Ribena Yu for financial support.


\begin{references}

\bibitem{tHooft} G. 't Hooft, Nucl. Phys. {\bf B72} (1974) 461.

\bibitem{z1} M. Bershadsky, Z. Kakushadze and C. Vafa, Nucl. Phys. {\bf B523}
(1998) 59.

\bibitem{z2} Z. Kakushadze, Nucl. Phys. {\bf B529} (1998) 157;  
Phys. Rev. {\bf D58} (1998) 106003; Phys. Rev. {\bf D59} (1999) 045007; 
Nucl. Phys. {\bf B544} (1999) 265.

\bibitem{malda} J.M. Maldacena, Adv. Theor. Math. Phys. {\bf 2} (1998) 231.

\bibitem{GKP} S.S. Gubser, I.R. Klebanov and A.M. Polyakov, Phys. Lett. 
{\bf B428} (1998) 105.

\bibitem{Witten} E. Witten, Adv. Theor. Math. Phys. {\bf 2} (1998) 253.

\bibitem{Kachru} S. Kachru and E. Silverstein, Phys. Rev. Lett. {\bf 80} 
(1998) 4855.

\bibitem{radu} Z. Kakushadze and R. Roiban, JHEP {\bf 0103} (2001) 043.

\bibitem{rozleigh} R.G. Leigh, M. Rozali, Phys. Rev. {\bf D59} (1999) 026004. 

\bibitem{dougmoore} M. Douglas and G. Moore, hep-th/9603167.

\bibitem{LNV} A. Lawrence, N. Nekrasov and C. Vafa, Nucl. Phys. {\bf B533} 
(1998) 199.

\bibitem{HSU} A. Hanany, M.J. Strassler and A.M. Uranga, JHEP {\bf 9806}
(1998) 011.

\bibitem{IRU} L.E. Ibanez, R. Rabadan and A.M. Uranga, 
Nucl. Phys. {\bf B542} (1999) 112.

\bibitem{Poppitz} E. Poppitz, Nucl. Phys. {\bf B542} (1999) 31.

\bibitem{KT} Z. Kakushadze and T.R. Taylor, Nucl. Phys. {\bf B562} (1999) 78.

\bibitem{early} 
V. Rubakov and M. Shaposhnikov, Phys. Lett. {\bf B125} (1983) 136.

\bibitem{BK}
A. Barnaveli and O. Kancheli, Sov. J. Nucl. Phys. {\bf 52} (1990) 576.

\bibitem{pol} J. Polchinski, Phys. Rev. Lett. {\bf 75} (1995) 4724.

\bibitem{witt} P. Ho{\u r}ava and E. Witten, Nucl. Phys. {\bf B460} (1996)
506; Nucl. Phys. {\bf B475} (1996) 94;\\
E. Witten, Nucl. Phys. {\bf B471} (1996) 135.

\bibitem{lyk} I. Antoniadis, Phys. Lett. {\bf B246} (1990) 377;\\
J. Lykken, Phys. Rev. {\bf D54} (1996) 3693.

\bibitem{shif} G. Dvali and M. Shifman, Nucl. Phys. {\bf B504} (1997) 127;
Phys. Lett. {\bf B396} (1997) 64.

\bibitem{TeV} N. Arkani-Hamed, S. Dimopoulos and G. Dvali, 
Phys. Lett. {\bf B429} (1998) 263; Phys. Rev. {\bf D59} (1999) 086004.

\bibitem{dien} K.R. Dienes, E. Dudas and T. Gherghetta, Phys. Lett. 
{\bf B436} (1998) 55; Nucl. Phys. {\bf B537} (1999) 47; hep-ph/9807522;\\
Z. Kakushadze, Nucl. Phys. {\bf B548} (1999) 205; Nucl. Phys.
{\bf B552} (1999) 3; Nucl. Phys. {\bf B551} (1999) 549.\\
Z. Kakushadze and T.R. Taylor, Nucl. Phys. {\bf B562} (1999) 78.

\bibitem{3gen} Z. Kakushadze, Phys. Lett. {\bf B434} (1998) 269; 
Nucl. Phys. {\bf B535} (1998) 311; Phys. Rev. {\bf D58} (1998) 101901.

\bibitem{anto} I. Antoniadis, N. Arkani-Hamed, S. Dimopoulos and G. Dvali,
Phys. Lett. {\bf B436} (1998) 257.

\bibitem{ST} G. Shiu and S.-H.H. Tye, Phys. Rev. {\bf D58} (1998) 106007.

\bibitem{BW} Z. Kakushadze and S.-H.H. Tye, Nucl. Phys. {\bf B548} (1999) 180;
Phys. Rev. {\bf D58} (1998) 126001.

\bibitem{Gog} M. Gogberashvili, hep-ph/9812296; Europhys. Lett. {\bf 49} 
(2000) 396.

\bibitem{RS} L. Randall and R. Sundrum, Phys. Rev. Lett. {\bf 83} (1999)
3370; Phys. Rev. Lett. {\bf 83} (1999) 4690.

\bibitem{DGP} G. Dvali, G. Gabadadze and M. Porrati, Phys. Lett. {\bf 
B485} (2000) 208.

\bibitem{DG} G. Dvali and G. Gabadadze, Phys. Rev. {\bf D63} (2001) 065007.

\bibitem{IK} A. Iglesias and Z. Kakushadze, hep-th/0011111; hep-th/0012049.

\bibitem{DGP1} G. Dvali, G. Gabadadze and M. Porrati, Phys. Lett. {\bf 
B484} (2000) 112; Phys. Lett. {\bf B484} (2000) 129.

\bibitem{witten} E. Witten, hep-ph/0002297.

\bibitem{zura} Z. Kakushadze, Phys. Lett. {\bf B488} (2000) 402;
Phys. Lett. {\bf B489} (2000) 207; Phys. Lett. {\bf B491} (2000) 317;
Mod. Phys. Lett. {\bf A15} (2000) 1879.

\bibitem{dvali} G. Dvali, hep-ph/0004057.

\bibitem{COSM} Z. Kakushadze, Nucl. Phys. {\bf B589} (2000) 75;
Phys. Lett. {\bf B497} (2001) 125;\\
O. Corradini and Z. Kakushadze, Phys. Lett. {\bf B494} (2000) 302;
Phys. Lett. {\bf B506} (2001) 167;\\
Z. Kakushadze and P. Langfelder, Mod. Phys. Lett. {\bf A15} (2000) 2265.

\bibitem{dilaton} G. Dvali and Z. Kakushadze, Phys. Lett. {\bf B417}
(1998) 50.

\bibitem{KaSh} Z. Kakushadze, Nucl. Phys. {\bf B512} (1998) 221;\\
Z. Kakushadze and G. Shiu, Phys. Rev. {\bf D56} (1997) 3686;
Nucl. Phys. {\bf B520} (1998) 75.

\bibitem{KST}Z. Kakushadze, G. Shiu and S.-H.H. Tye,
Nucl. Phys. {\bf B533} (1998) 25; Phys. Rev. {\bf D58} (1998) 086001. 

\bibitem{NPO} Z. Kakushadze, 
Phys. Lett. {\bf B455} (1999) 120; Int. J. Mod. Phys. {\bf A15} (2000) 3461;
Phys. Lett. {\bf B459} (1999) 497; Int. J. Mod. Phys. {\bf A15} (2000) 3113.

\end{references}
\end{document}